\documentclass[aps,prr,twocolumn,amsfonts, amssymb, amsmath,showpacs,superscriptaddress,showkeys, nofootinbib,longbibliography]{revtex4-1}
\AtBeginDocument{%
    \newwrite\bibnotes
    \def\bibnotesext{Notes.bib}
    \immediate\openout\bibnotes=\jobname\bibnotesext
    \immediate\write\bibnotes{@CONTROL{REVTEX41Control}}
    \immediate\write\bibnotes{@CONTROL{%
    apsrev41Control,author="08",editor="1",pages="1",title="0",year="1"}}
     \if@filesw
     \immediate\write\@auxout{\string\citation{apsrev41Control}}%
    \fi
}%
\usepackage[T1]{fontenc}
\usepackage[english]{babel}
\usepackage{siunitx}
\usepackage[utf8]{inputenc}
\usepackage{url}
\usepackage{dsfont}
\usepackage{bbm}
\usepackage[usenames,dvipsnames]{xcolor}
\usepackage{nicefrac}
\usepackage{setspace}
\usepackage[short]{ optidef }
\usepackage{float}
\usepackage{comment}

\usepackage[colorlinks=true, citecolor=blue,urlcolor=blue,linkcolor=blue,filecolor=black]{hyperref}

\usepackage[colorinlistoftodos, color=green!40, prependcaption]{todonotes}
\usepackage{ulem}
\usepackage{amsthm}
\usepackage{mathtools}
\usepackage{physics}
\usepackage{xcolor}
\usepackage{graphicx}
\usepackage{adjustbox}
\usepackage{placeins}

\usepackage{lipsum}
\usepackage{csquotes}

\newcommand{\beq}{\begin{equation}}
\newcommand{\eeq}{\end{equation}}
\renewcommand{\emph}{\textit}
\renewcommand{\tr}{\text{Tr}}
\usepackage{color,soul}

\definecolor{pinocol}{rgb}{.1,0.7,0.1}

\newcommand{\marco}[2]{{\color{blue}\sout{#1}#2}}

\begin{document}
    
    \title{
    Semi-device independent randomness from 
    $d$-outcome continuous-variable detection}


    \author{Hamid Tebyanian}
    \affiliation{
    	Dipartimento di Ingegneria dell'Informazione, Universit\`a di Padova, via Gradenigo 6B, 35131 Padova, Italy}
    \author{Marco Avesani}
    \affiliation{
    	Dipartimento di Ingegneria dell'Informazione, Universit\`a di Padova, via Gradenigo 6B, 35131 Padova, Italy}
    \author{Giuseppe Vallone}
    \affiliation{
    	Dipartimento di Ingegneria dell'Informazione, Universit\`a di Padova, via Gradenigo 6B, 35131 Padova, Italy}
    \affiliation{
    	Dipartimento di Fisica e Astronomia, Universit\`a di Padova, via Marzolo 8, 35131 Padova, Italy}
    \affiliation{Istituto di Fotonica e Nanotecnologie - CNR, Via Trasea 7 - 35131 Padova, Italy}

    \author{Paolo Villoresi}
    \affiliation{
    	Dipartimento di Ingegneria dell'Informazione, Universit\`a di Padova, via Gradenigo 6B, 35131 Padova, Italy}
    \affiliation{Istituto di Fotonica e Nanotecnologie - CNR, Via Trasea 7 - 35131 Padova, Italy}

    \begin{abstract}
    Recently, semi-device independent protocols have attracted increasing attention, guaranteeing security with few hypotheses and experimental simplicity. In this paper, we demonstrate a many-outcomes scheme with the binary phase-shift keying (BPSK) for a semi-device independent protocol based on the energy assumption. We show in theory that the number of certified random bits of the $d$-outcomes system outperforms the standard scheme (binary-outcomes). Furthermore, we compare the results of two well-known measurement schemes, homodyne and heterodyne detection.  Lastly, taking into account the experimental imperfections, we discuss the experimental feasibility of the $d$-outcome design.
    \end{abstract}
    
    \maketitle
    
    \section{Introduction}
    In the information security age, data privacy and secure communication are of paramount relevance. It is worth to stress the role of genuine random numbers for privacy and security applications. Nearly all of the protocols dealing with privacy and security relies on random numbers, and the protocol's security is directly connected to the quality of the employed random numbers~\cite{Stipcevic12}. Thus, owning certified random numbers is a critical component for guarding the information. Pseudo-random number generators have been popular and widely used in the past few decades. However, the generated numbers are not truly random since the randomness source is based upon a classical phenomenon that is deterministic. In general, random number generators (RNG) can be classified into two major groups, classical and quantum. 
    Due to their determinism, Classical RNG cannot offer high levels of security, while quantum random number generators (QRNG), are qualified candidates for generating genuine and unpredictable random numbers based on the intrinsic randomness of quantum mechanics~\cite{Herrero-Collantes2017}.

    Despite the fact that quantum mechanics assures the unpredictability of the generated random numbers, experimental imperfections of QRNG can open a 
    backdoor for eavesdroppers to attack or manipulate the protocol~\cite{Acin2016}. For instance, the generator's apparatus can be correlated with an external party, or deviate from the expected behaviour. Hence QRNGs can be categorized into three subgroups, trusted-device, semi-device independent (semi-DI), and device-independent (DI) QRNGs \cite{Ma2016}.
    Although the trusted-deceive QRNGs are cheap, fast, and more reliable than the classical generators, they can be compromised due to the security loopholes resulting from trusting the devices. On the other hand, the highest security is achievable by DI QRNGs where randomness is certified by the violation of a Bell inequality, without any trust on any devices~\cite{Liu2018a}. 

    Besides offering highly secure randomness, it also allows the devices to be undependable and, hence, robust against experimental imperfections.
    Unfortunately, the experimental realization of a loophole-free Bell test is extremely hard to accomplish, and only proof-of-principle experiments were realized, obtaining modest generation rates ~\cite{Bierhorst2017,Acin17,Liu2019b, Ma_new2,Zhang2020}.
    Taking into account the complexity of this protocol and the low bit rate, the DI QRNGs are still very far from being practical. 
    Indeed, security and speed are the two key features of RNG and both are needed in practical applications.
    
    Semi-DI protocols are an intermediate approach between DI and trusted-device schemes, which offer an optimal trade-off between generation rate, security, and ease-of-implementation~\cite{Ma2016}.
    Depending on the protocol needs, assumptions can vary; for very secure protocols, there are fewer assumptions on the device, i.e., single assumption on the overlap or energy of the prepared states~\cite{Brask2017,VanHimbeeck2017semidevice,Himbeeck2019CorrelationsConstraints,Rusca2019,Rusca2020,avesani2020}. 
    Depending on the protocol needs, assumptions can vary; 
    they can be related to the dimension of Hilbert space~\cite{Lunghi2014b, Canas2014}, they may require trusted measurement in the case of source-DI protocols \cite{Avesani2018,Ma_new1,Marangon} or they may assume a trusted source, an in measurement-DI protocols ~\cite{Nie2016, Cao2015}. Recently a new class of protocols has been proposed, where both source and measurement are untrusted and only a single assumption on the overlap or energy of the prepared states is required~\cite{Brask2017,VanHimbeeck2017semidevice,Himbeeck2019CorrelationsConstraints,Rusca2019,Rusca2020,avesani2020}.
    These protocols can provide an increased security, since they reduce the number of assumptions on the devices.

    In this work, we investigate the impact of increasing the number of outcomes of the measurement apparatus given a binary-input semi-DI QRNG~\cite{avesani2020,Rusca2020}.
    The protocol builds upon the prepare-and-measure scheme, with a single assumption on the maximum energy of the prepared states that implies a lower bound on the state's overlap. The implementation is based on optical continuous variables (CVs), that allow high generation rates.
    
    As shown in~\cite{Mari2019}, for a SDI-QRNG with $n$ inputs subjected to the overlap bound, the  measurement apparatus achieving the maximum randomness is obtained by using an $n+1$ outcome POVM and no more than $\log_2 (n + 1)$ random bits can actually be certified. 
    However, such optimal POVM is not easily obtained with typical CV measurements and we will show that, if the measurement is realized by using homodyne or heterodyne detector, increasing the number of outputs to more than 3 (for 2 inputs) will improve the generation rate.
    In particular, we will report the numerical results of the method employed for randomness estimation, from three to fourteen outcomes, concerning both homodyne and heterodyne detections and then compare it with the binary outcomes result.
    We will also investigate the generation rate as a function of the system efficiency, showing that the advantage of increasing the number of outcomes decreases with lower efficiency.

    \begin{figure}[t!]
    \includegraphics[width=\linewidth]{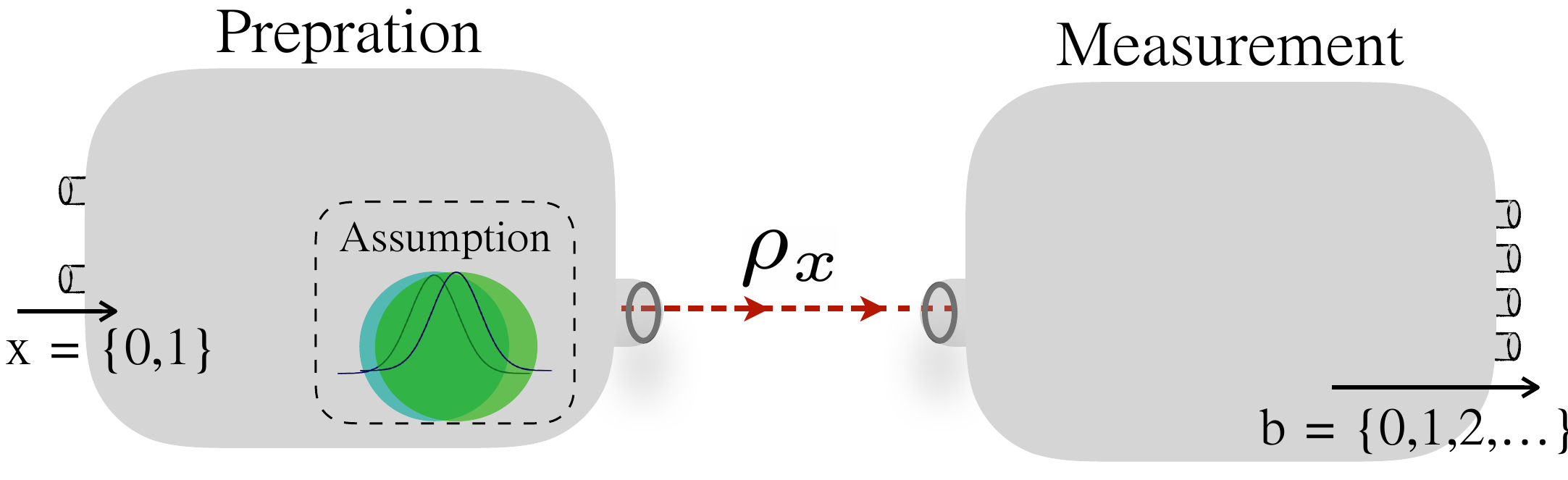}
    \caption{The general design of QRNG protocol. Depending on the input $x$, the unknown state $\rho_0$ or $\rho_1$ is transmitted from the preparation part. A single assumption is present on the state's energy. The measurement device, with no assumptions, performs a generic measurement and outputs $b\in\{0,\dots,d-1\}$.}
    \label{fig:ExperimentalScheme1}
    \end{figure} 

    \section{SDI-QRNG model}
    \subsection{General framework}
    The protocol is based on two untrusted devices, the preparation and measurement, and a single assumption corresponding to an upper bound on the prepared state's energy. Similar approaches were presented in \cite{Brask2017,avesani2020,Himbeeck2019CorrelationsConstraints}. A general scheme of this protocol is shown in Fig. \ref{fig:ExperimentalScheme1}:

    a preparation device emits the unknown states $\rho_x$ after receiving the binary input $x \in \{ 0, 1 \}$ from the user. The measurement device has $d$ outputs $b \in \{0,1,\cdots,d-1\} $.
    By running $N$ times the experiment it is possible to estimate the conditional probabilities $p(b|x)$.

    The measurement device is considered as a black box, whose internal working principles are unknown to the user. 
    The preparation device is a `gray box'': 
    the internal working principles are not known but it comes with an assumption, namely an upper bound on the energy of the prepared states
    \beq
    \label{eq:energy_bound}
    \langle \hat n\rangle_{\rho_x}\leq \mu\,.
    \eeq

    As shown in \cite{Tomamichel2011}, the conditional min-entropy, namely the amount of genuine random bits per measurement run is given by

    \beq
    \label{eq:Hmin}
    H_{\rm min}=-\log_2\left(P_{\rm g}\right)
    \eeq
    where $P_{\rm g}$ is the guessing probability, namely the highest probability that an attacker knowing the internal working principle of the devices can guess the outcomes $b$, given the input $x$. 
    It is worth to note that the bound on the energy, whose validity can be checked experimentally, implies a lower bound on the scalar product between the emitted states~\cite{Himbeeck2019CorrelationsConstraints,avesani2020}
    and thus the approach of \cite{Brask2017} can be followed to obtain $P_{\rm g}$ from the experimental data.

    By generalizing  the approach of \cite{Brask2017} with $d$ outcomes, $P_{\rm g}$ can be found as the solution of the following
    semidefinite programming (SDP)
    
    \begin{maxi}|l|
      {
      M_b^{\lambda_0,\lambda_1}
      }
      {\widetilde P_{\rm g}=
      \frac{1}{2}\sum\limits_{x = 0}^1 
      \sum\limits_{{\lambda _0},{\lambda _1} = 0}^{d-1} \bra{\psi_x}
      M_{\lambda_x}^{{\lambda _0},{\lambda _1}}
       \ket{\psi_x}
      }
      {}{}
      \addConstraint{M_b^{{\lambda _0},{\lambda _1}} = {(M_b^{{\lambda _0},{\lambda _1}})^\dag }}
      \addConstraint{M_b^{{\lambda _0},{\lambda _1}} \ge 0}
      \addConstraint{\sum_{b=0}^{d-1} {M_b^{{\lambda _0},{\lambda _1}}}  = \frac12 \tr[\sum_{b=0}^{d-1} {M_b^{{\lambda _0},{\lambda _1}}} ]\mathbb{I}}
      \addConstraint{\sum\limits_{{\lambda _0},{\lambda _1} = 0}^{d-1} 
      \bra{\psi_x}
      M_b^{{\lambda _0},{\lambda _1}}
      \ket{\psi_x}
      = p(b|x)\,,\quad \forall b,x}
      \label{eq:SDP_primal}
    \end{maxi}
    where $M_b^{{\lambda _0},{\lambda _1}}$ are $2\times 2$ operators in the 2-dimensional Hilbert space spanned by the orthonormal vectors $\ket{0}$ and $\ket{1}$ and the states $\ket{\psi_x}$ are defined by
    \begin{equation}
    \begin{aligned}
        \ket{\psi_0}&=\ket0\,,
        \\
        \ket{\psi_1}&=(1-2\mu)\ket0+
        2\sqrt{\mu(1-\mu)}\ket1\,.
    \end{aligned}
    \label{state:Pino}
    \end{equation}
  
    The above states $\ket{\psi_x}$ saturates the bound
    $|\braket{\psi_0}{\psi_1}|\geq 1-2\mu$ derived from \eqref{eq:energy_bound}, and can be used in the optimization without loss of generality see \cite{VanHimbeeck2017semidevice,avesani2020}. 
    In Eq. \eqref{eq:SDP_primal} we assumed that the input states
    are prepared with equal probability, namely $p_x=1/2$.

    The variables $\lambda\equiv(\lambda_0,\lambda_1)$ represent the classical information available to anyone knowing the internal working of the device. The operators $M_b^{\lambda_0\,\lambda_1}$ are related to possible physical realizations of the measurement device that are compatible with the observed probabilities $p(b|x)$.
    More precisely, for each value of the pair $(\lambda_0\,\lambda_1)$, the value
    $q_\lambda=\frac12\tr[\sum_{b} {M_b^{{\lambda _0},{\lambda _1}}} ]$ represents the probability that the measurement device is actually implementing the POVM defined by the operators
    $\{\Pi_b^{\lambda_0\,\lambda_1}\}$ where $\Pi_b^{\lambda_0\,\lambda_1}
    =M_b^{\lambda_0\,\lambda_1}/q_\lambda$.

    It is worth noticing that the above approach is general and does not depends on the actual implementation of the preparation and measurement devices. The min-entropy is directly calculated by using only the value of the energy bound $\mu$ and the measured output probabilities $p(b|x)$, independently of their physical realization. We observe that larger $H_{\rm min}$
    can be obtained whenever the probabilities $p(b|x)$
    allow to better distinguish the two input states.

    \begin{figure}[h!]
    \includegraphics[width=\linewidth]{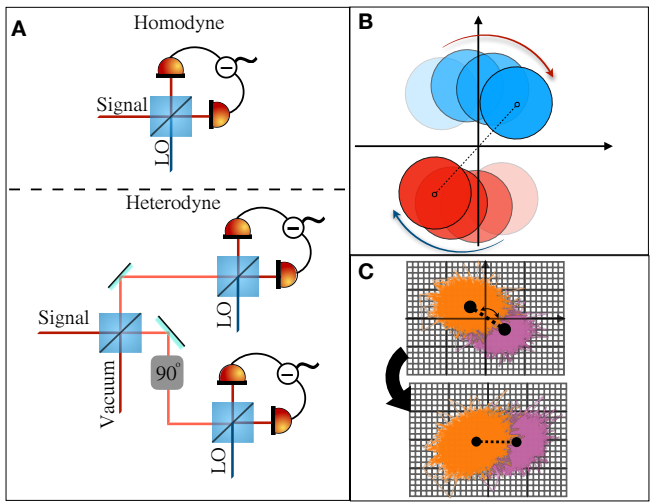}
	\caption{Homodyne and heterodyne detection. a) Representation of the two detection schemes, b) Effects of the phase instability on the received states, c) Offline phase compensation for heterodyne detection.}
	\label{fig:het_hom}
    \end{figure}

    \subsection{Implementation with continuous variables}
    We now illustrate the amount of randomness that can be obtained by using single-mode optical continuous variables defined by
    the creation operator $\hat a^\dag$.
    \subsubsection{Preparation}

    In the preparation part, we employed the Binary Phase Shift-Keying (BPSK) system, where the source, a continuous-wave (CW) laser, emits two coherent states with the same mean-photon number and a $\pi$ phase shift $\ket{\psi_0}=\ket{\alpha}$ and $\ket{\psi_1}=\ket{-\alpha}$. We can use the representation of a coherent state in Fock space to define $\ket{\alpha}$ as  
    $\ket{\pm\alpha}=
    {e^{-\frac{\mu}{2}}}\sum\limits_{n = 0}^\infty  \frac{(\pm\sqrt{\mu}e^{i\phi})^n}{{\sqrt{n!}}}\ket n$, where $\alpha=\sqrt{\mu}e^{i\phi}$, $\mu$ is the mean photon number and $\phi$ is the relative phase between the signal and the local oscillator (LO). We here assume that the LO is chosen such that $\phi=0$.
    Note that the input $x$ should be uncorrelated with $\lambda$ and independent of the devices. Thus they can be generated from a standard RNG (e.g., Pseudo RNG). 
    We note that the mean photon number for each state $\ket{\psi}$ is upper-bounded by the quantity $\mu$ given in Eq. \eqref{eq:energy_bound}. We note that states with non-vanishing overlap cannot be deterministically distinguished, unlike orthogonal states.

    \subsubsection{Measurement}
    Homodyne and heterodyne tomography are two primary and well-established detection schemes for measuring CV states of light, see Fig.~\ref{fig:het_hom}. By homodyning, the quantum state is measured from samples obtained from projected Wigner functions, whereas heterodyne detection directly samples phase space coordinates from the Husimi Q‐function~\cite{Muler2016,Walker07}.
    For what regards Semi-DI QRNG protocols, both heterodyne and homodyne detection have been employed at the receiver side, as shown in \cite{avesani2020} and \cite{Rusca2020}, respectively.  In these works, the (potentially) infinite outcomes of the CV measurement are grouped into two disjoint sets, corresponding to a binary outcome. Here we consider the more general case in which the physical outcomes can be grouped into a larger number of sets.
    
    The POVM of homodyne and heterodyne receivers can be represented respectively by\marco{}{:}
    \begin{equation}
    \begin{aligned}
    \Pi^{(\rm hom)}(X)&=\ket{X}\bra{X}
    \\
    \Pi^{(\rm het)}(\beta)&=\frac{1}{\pi}\ket{\beta}\bra{\beta}
    \end{aligned}
    \eeq
    where $\ket{X}$  is the eigenstate of the $\hat X=(\hat a+\hat a^\dag)/\sqrt{2}$ operator and $\ket{\beta}$ is the coherent state with complex amplitude $\beta$.

    The corresponding probability densities associated to the measurement of the states  $\ket{\pm\sqrt{\mu}}$ are given by
    \beq
    \label{prob_density}
    \begin{aligned}
    \mathcal P^{(\rm hom)}_\pm(X)&
    =\sqrt{\frac{2}{\pi}} e^{-2(X\mp\sqrt{\eta\mu})^2}\,,
\\
    \mathcal P^{(\rm het)}_\pm(\beta)&
    =\frac{1}{\pi } e^{-(X\mp\sqrt{\eta\mu})^2} e^{-Y^2}\,,
    \end{aligned}
    \eeq
    with real $X$, $Y$  and $\beta=X+iY$. In the above equations we included
    the overall efficiency $\eta$ of the channel and of the receiver devices.
    In order to obtain $d$ possible outcomes $b=0,1,\cdots d-1$ we need to partition the real line $(X)$ or the phase space ($\beta$) into $d$ disjoint sets.
    
    In the homodyne case, it is necessary to choose $d-1$ increasing real numbers $X_1<X_2<\ldots<X_{d-1}$ such that the outcome probabilities for $b=0,\cdots,d-1$ can be written as
    \beq
    \label{prob_homo}
    \begin{aligned}
    p^{(\rm hom)}(b|x)
    &=\frac{1}{\sqrt{\pi}} \int_{X_b}^{X_{b+1}} e^{-(X-(-1)^x\sqrt{2\eta\mu})^2} dX
    \\
    &=\frac12\left[
    \erf(X_{b+1}-(-1)^x\sqrt{2\eta\mu})\right.
    \\
    &\qquad\quad \left.-\erf(X_{b}-(-1)^x\sqrt{2\eta\mu})\right]
    \end{aligned}
    \eeq
    with the convention that $X_0=-\infty$ and $X_d=+\infty$.
    We note that from Eq. \eqref{prob_density} to Eq. \eqref{prob_homo}
    we have performed a change of the integration variable.

    In the heterodyne case, we may define a partition of the phase space $\{\Lambda_b\}$ with $d$ elements. The output probabilities can be written as
    \beq
    \begin{aligned}
    p^{(\rm het)}(b|x)&=\frac{1}{\pi } \int_{\Lambda_b} e^{-(X-(-1)^x\sqrt{\eta\mu})^2} e^{-Y^2}
    dXdY
    \\
    &=\frac{e^{-\eta\mu}}{\pi } \int_{\Lambda_b} re^{-r^2+2r(-1)^x\sqrt{\eta\mu}\cos\theta}
    drd\theta
    \end{aligned}
    \label{het_povm}
    \eeq
    
    In the following we will analyse the achievable randomness by considering the above measurements. We will consider the cases with an increasing number of outcomes and we compare it with the results obtained with 2 outcomes and already reported in \cite{avesani2020,Rusca2020}.

    \section{Results}
    \subsection{Homodyne detection}

    We start by considering the Homodyne detection with perfect efficiency ($\eta=1$). Due to the symmetry of the prepared states, the partition of the real axis is optimal when is symmetric around the origin. For instance, the configuration corresponding to 2, 3, 4 and 6 outcome are shown in table \ref{tab:homo_configuration} and are illustrated in Fig. \ref{fig:conf1}.
    \begin{table}[h!!!!]
        \centering
        \begin{tabular}{|c|c|c|c|c|c|c|c|}
             \hline
             outcomes  & $X_0$ & $X_1$ & $X_2$ & $X_3$ & $X_4$ & $X_5$ & $X_6$
             \\\hline
             2 & $-\infty$ & $0$ & $+\infty$ & / & / & / & /
             \\
             3 & $-\infty$ & $-L_1$ & $+L_1$ & $+\infty$ & / & / & /
             \\
             4 & $-\infty$ & $-L_1$ & $0$ & $+L_1$ & $+\infty$ & / & /
             \\
             6 & $-\infty$ & $-L_2$ & $-L_1$ & $0$ & $+L_1$ & $+L_2$
             & $+\infty$
             \\\hline
        \end{tabular}
        \caption{Definition of the partitions of the real axis corresponding to different output configurations for the homodyne detection.}
        \label{tab:homo_configuration}
    \end{table}

    \begin{figure}[h!]
    \includegraphics[width=\linewidth]{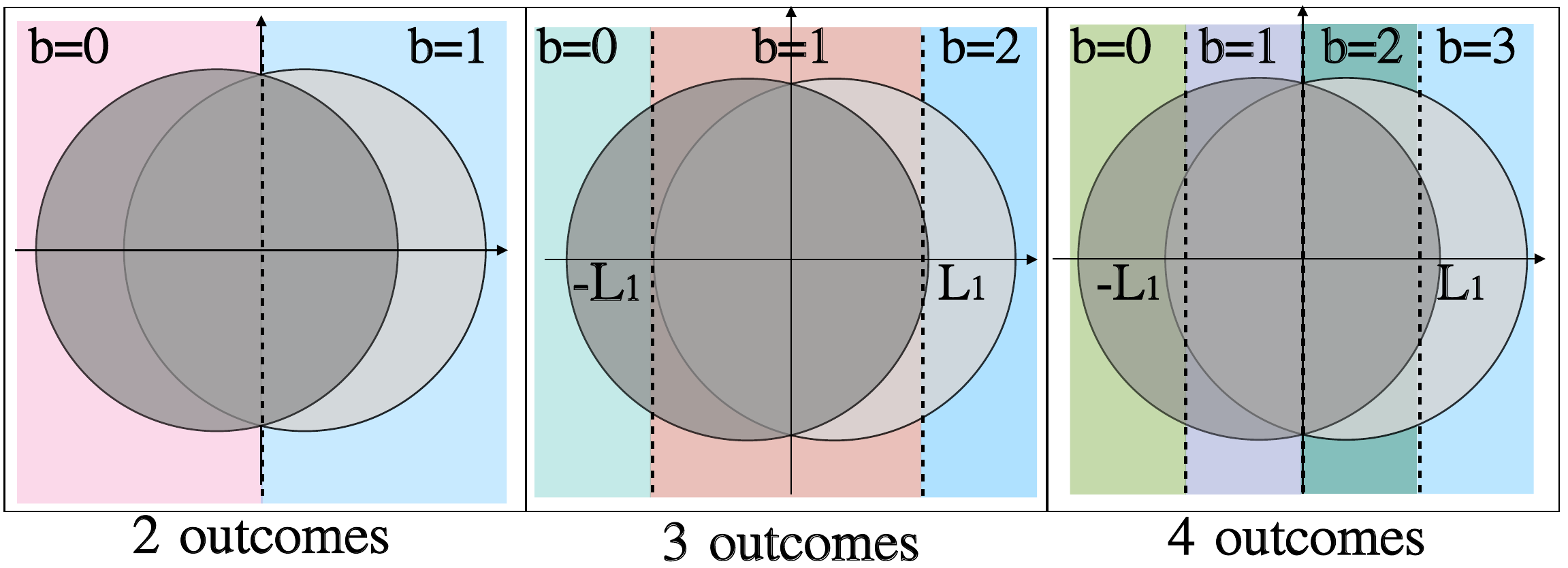}
	\caption{Homodyne measurement configurations.}
	\label{fig:conf1}
    \end{figure} 

    The amount of extractable genuine random bits is estimated by numerically solving the dual of the SDP optimization problem given by Eq. \eqref{eq:SDP_primal}, constrained by the conditional probabilities $p^{\rm hom}(b|x)$, obtained from Eq. \eqref{prob_homo}, together with the energy bound assumption $\mu$. The results are further optimized over the values $L_k$.
    The value of the min-entropy 
    as a function of the energy bound $\mu$ are shown in Fig. \ref{fig:homodyne} for the 2, 4 and 6 outcome cases.
    \begin{figure}[h!]
    \includegraphics[width=\linewidth]{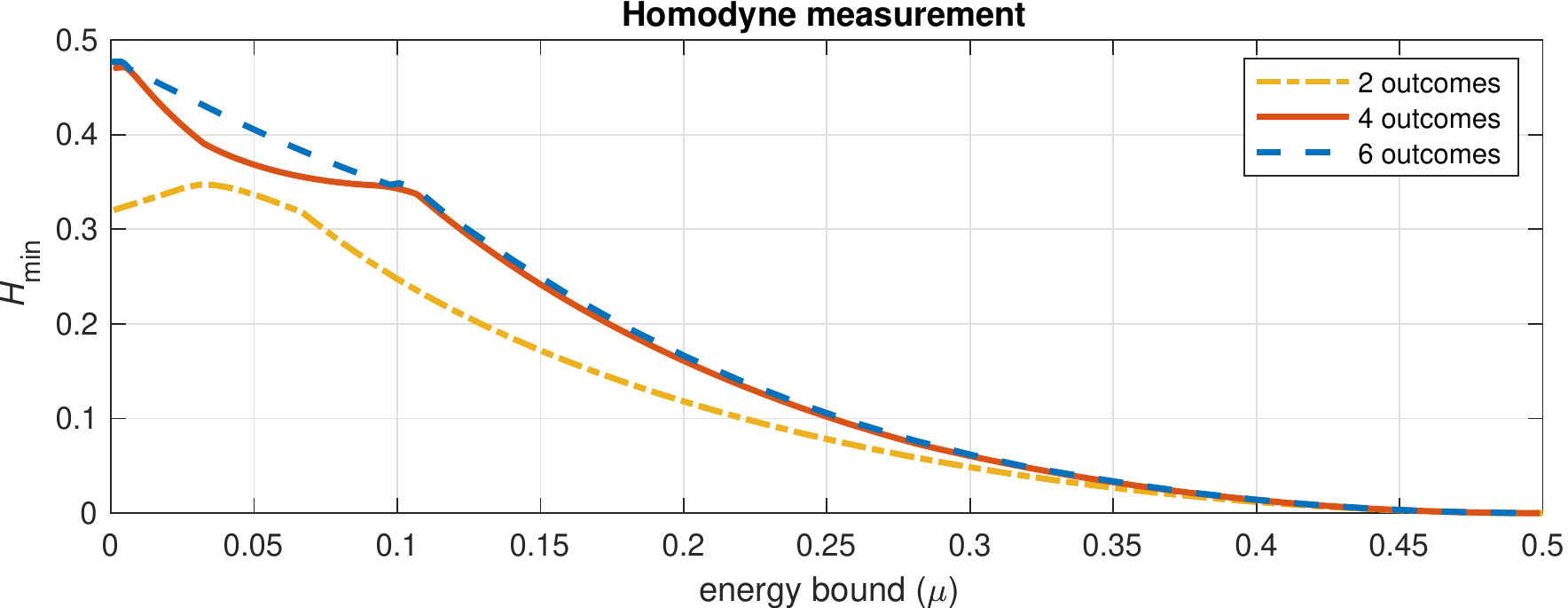}
	\caption{Min-entropy as a function of the energy bound  $\mu$ for homodyne detection and different numbers of outcomes.}
	\label{fig:homodyne}
    \end{figure} 
    
    \begin{figure}[h!]
    \includegraphics[width=\linewidth]{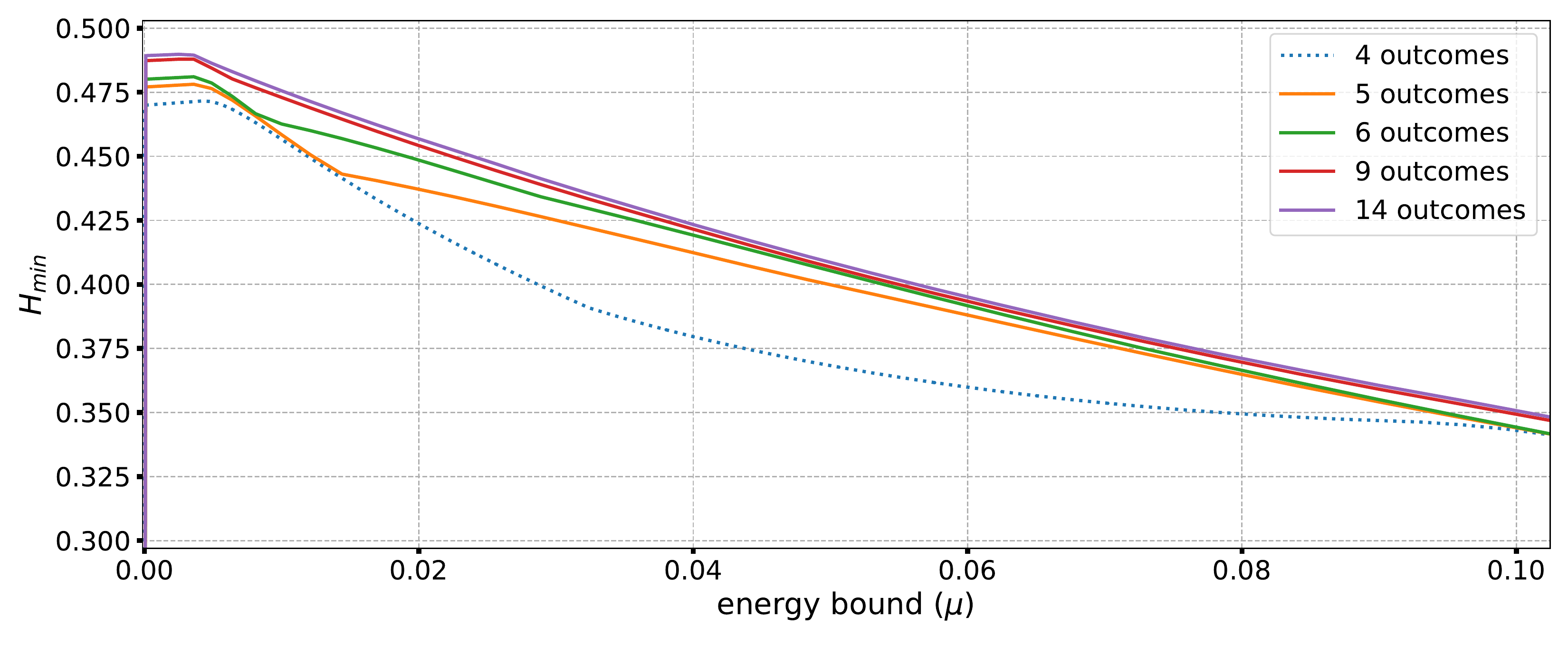}
	\caption{Min-entropy for large number of outcomes plotted for small $\mu$ values.} 
	\label{fig:nine}
    \end{figure}  

    As shown in  Fig. \ref{fig:homodyne}, by increasing the measurement outcomes the min-entropy monotonically increases over the entire range of $\mu$, meaning that
    more randomness can be certified. It is worth to note that, starting from the same physical implementation (homodyne measurement) and changing the post-processing (namely by changing the partitions of the outcomes) different values of the min-entropy can be obtained.
    
    One could ask what happens by further increasing the number of outcomes. As shown in Fig. \ref{fig:nine}, improvements are obtained for small values of $\mu$ by increasing the number of outcomes up to 14. In Fig. \ref{fig:more_outcomes} the
    best min-entropy (with optimized $\mu$) is shown in function of the number of outcomes. 
    {The data suggest that larger min-entropy will be obtained by further increasing the number of outcomes towards a seemingly asymptotic value of $0.5$.}

    \begin{figure}[h!]
    \includegraphics[width=\linewidth]{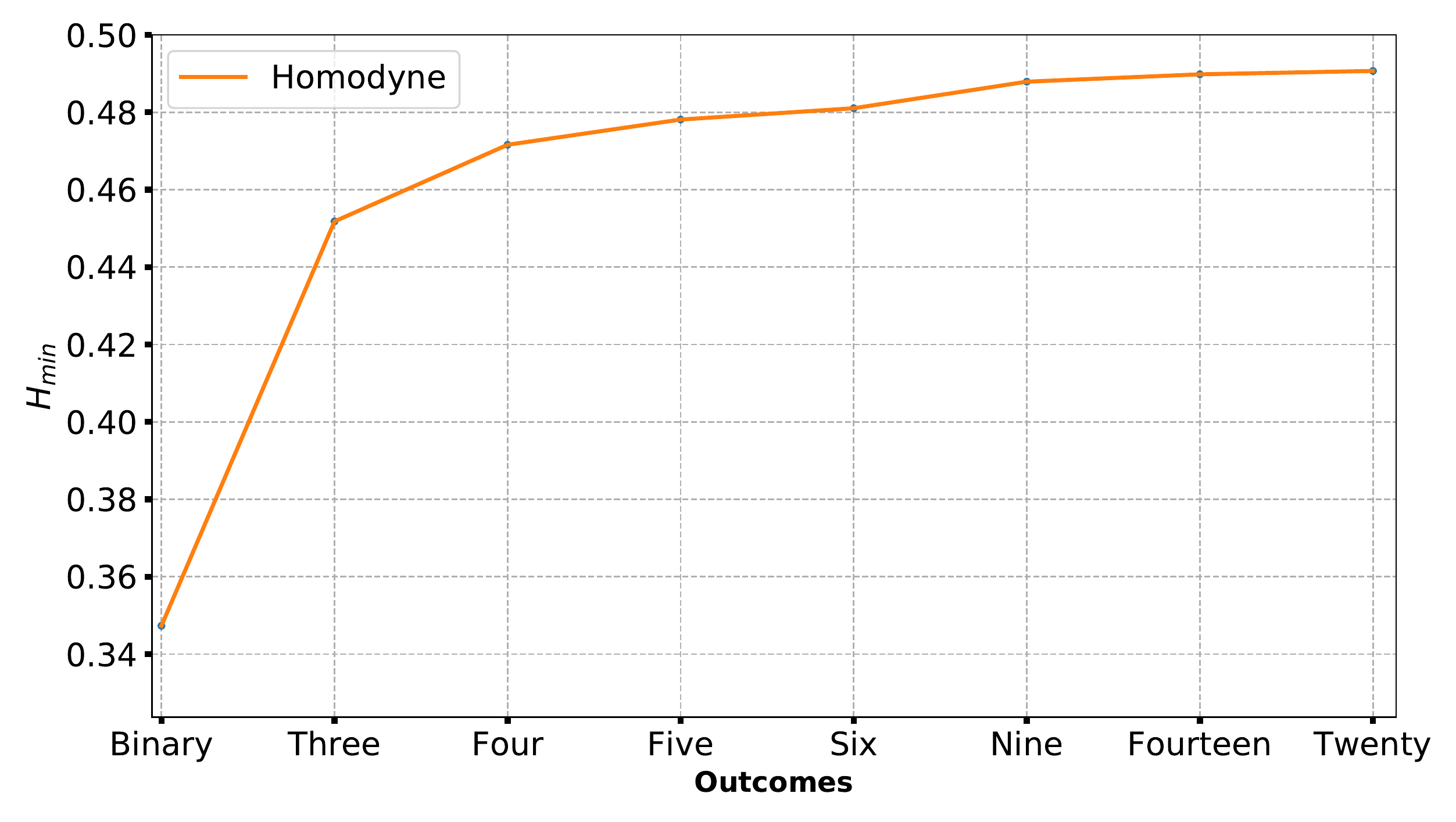}
	\caption{Maximum min-entropy (with optimized $\mu$ and  $\eta=1$) for the different number of outcomes for homodyne detection.} 
	\label{fig:more_outcomes}
    \end{figure}

    We now present the results obtained with inefficient system, namely by considering $\eta<1$. This parameter $\eta$ is used to model the effect of different experimental imperfections, such as the losses of the channel, the limited efficiency of the receiver's detectors or the electronic noise of the detection apparatus.  We carried out the same analysis described above by considering different values of $\eta$.
    We show in Fig. \ref{fig:homodyne} the min-entropy as a function of $\mu$ for different values of $\eta$ and for 2 and 4 outcomes.
    The corresponding optimal value of $L_1$ for the 4-outcome case are shown in Fig. \ref{fig:L}.
    From the figures, it can be shown that when the efficiency decreases, the advantage of using more outcomes is less evident, but it is still present.

    \begin{figure}[h!]
	\includegraphics[width=\linewidth]{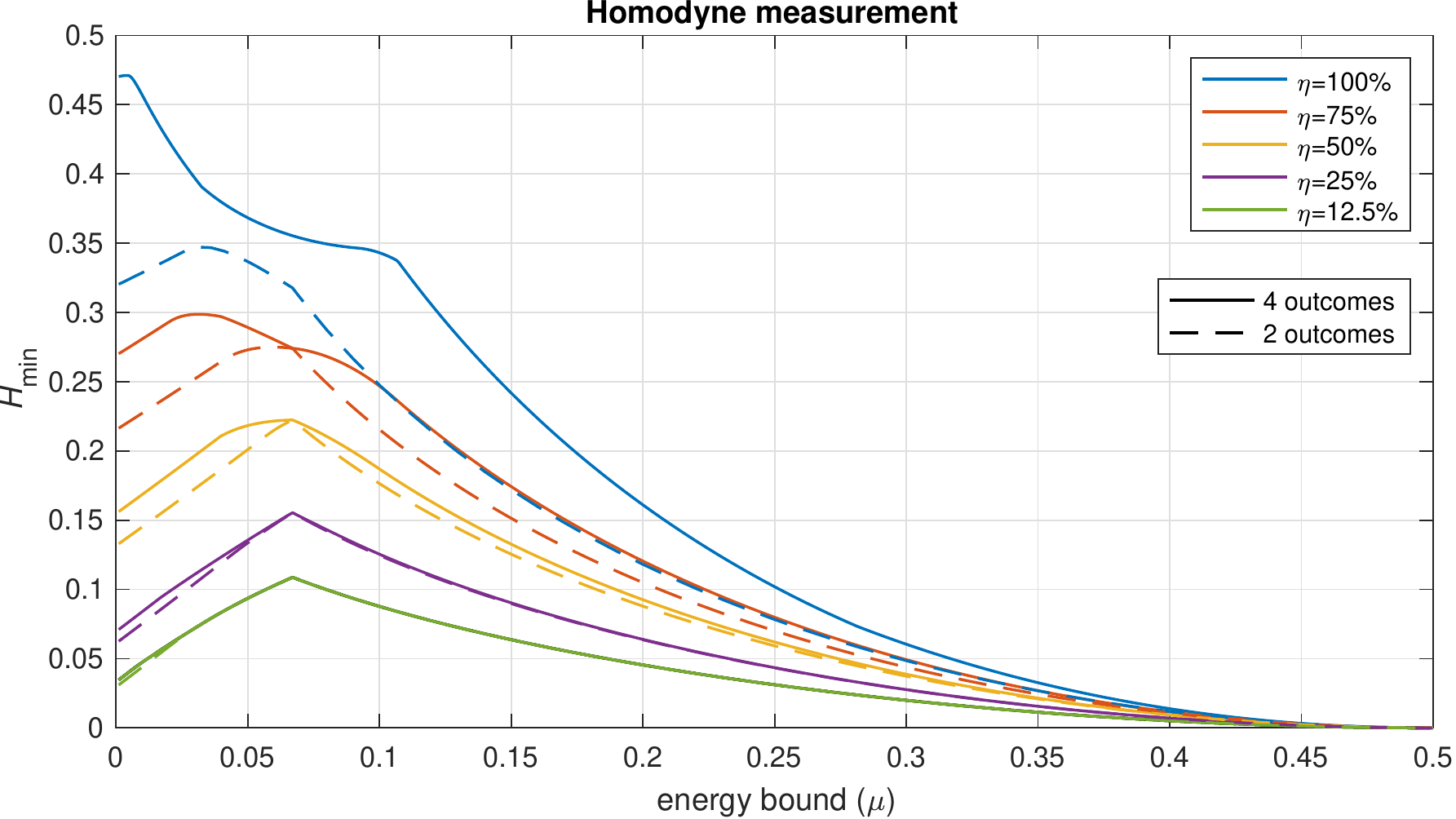}
	\caption{Min-entropy as a function of the energy bound $\mu$ for the homodyne detector. We compared  the 2-outcome (dashed line) and 4-outcome (solid line) scheme, for different values of the efficiency $\eta$.}
	\label{fig:homodyne}
    \end{figure}    
    
    \begin{figure}[h!]
    \includegraphics[width=\linewidth]{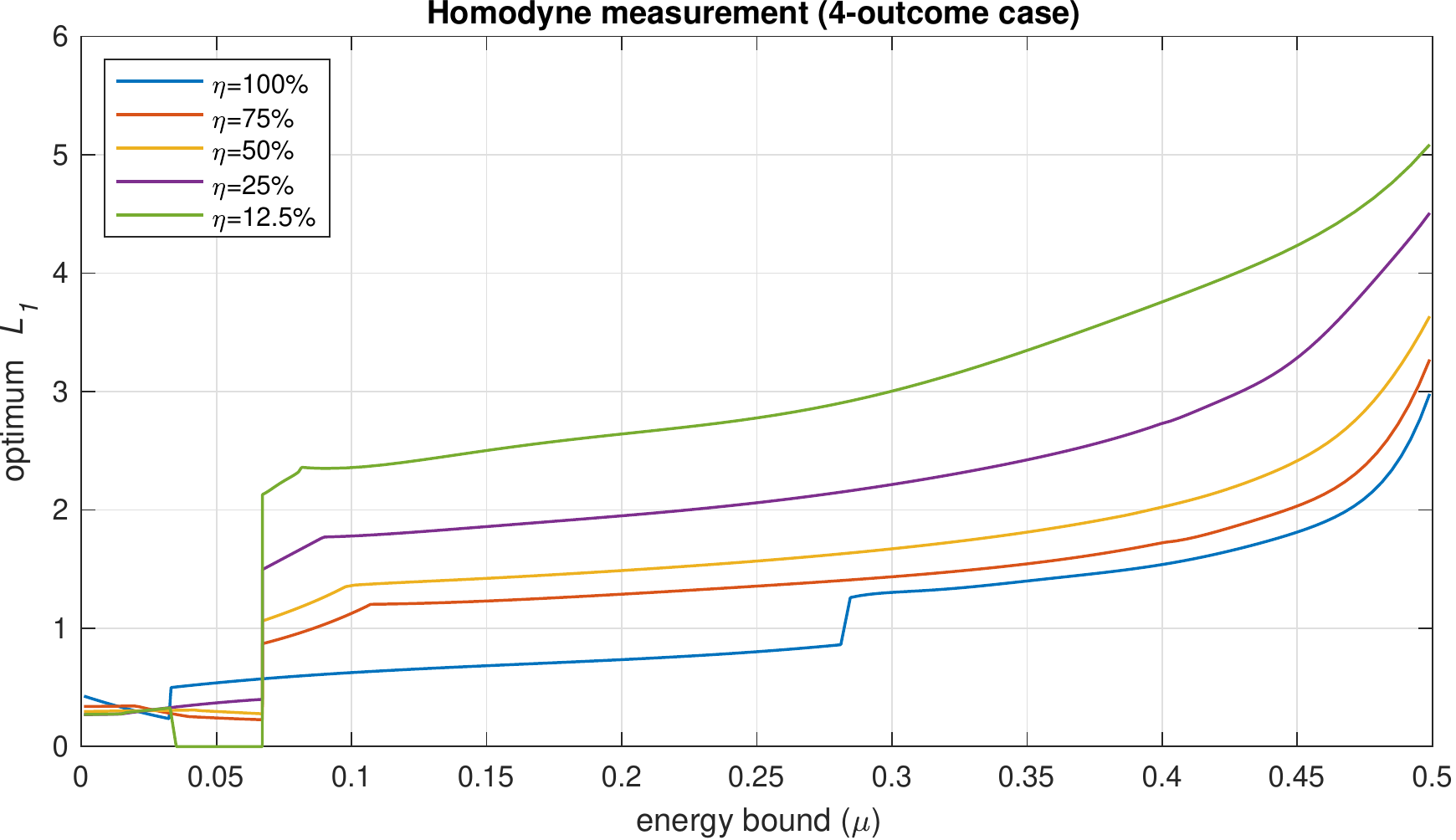}
	\caption{Optimal value of $L_1$ for the symmetric 4-outcome configuration for different system efficiency $\eta$.}
	\label{fig:L}
    \end{figure}

    \subsection{Heterodyne detection}
    Homodyne detection is only sensitive to one field quadrature, e.g.,  $X_{\phi}$ sampling only a projection of the phase space.  Heterodyne detection, on the other hands, performs a joint ``noisy'' measurement of two conjugated field quadratures, $\tilde X_{\phi}$ and $\tilde P_{\phi}$, thus sampling the entire phase-space. The number of possible (and potentially optimal) partitions for heterodyne detection is larger than homodyne, due to the increased dimensionality of the measurement.

    Similar to homodyne, it is possible to choose a ``strip'' partition, namely the configuration
    illustrated in Fig. \ref{fig:conf1}: the phase-space is subdivided in vertical strips whose boundaries are defined by the increasing real numbers
    $X_1<X_2<\ldots<X_{d-1}$. 
    Looking at Eq. \eqref{prob_homo} and \eqref{het_povm} it is possible to note that the heterodyne measurement with this configuration and efficiency $\eta$ is equivalent to the homodyne measurement with efficiency $\eta/2$.
    Thus, we can directly refer to Fig. \ref{fig:homodyne} for the results. 
    
    Other possible configurations are displayed in Fig. \ref{fig:conf2}. By running the SDP for all the configurations represented in Fig. \ref{fig:conf2}, we obtained a min-entropy that
    is always lower than the one obtained with the
    configuration shown in Fig. \ref{fig:conf1}. 
    \begin{figure}[h!]
    \includegraphics[width=\linewidth]{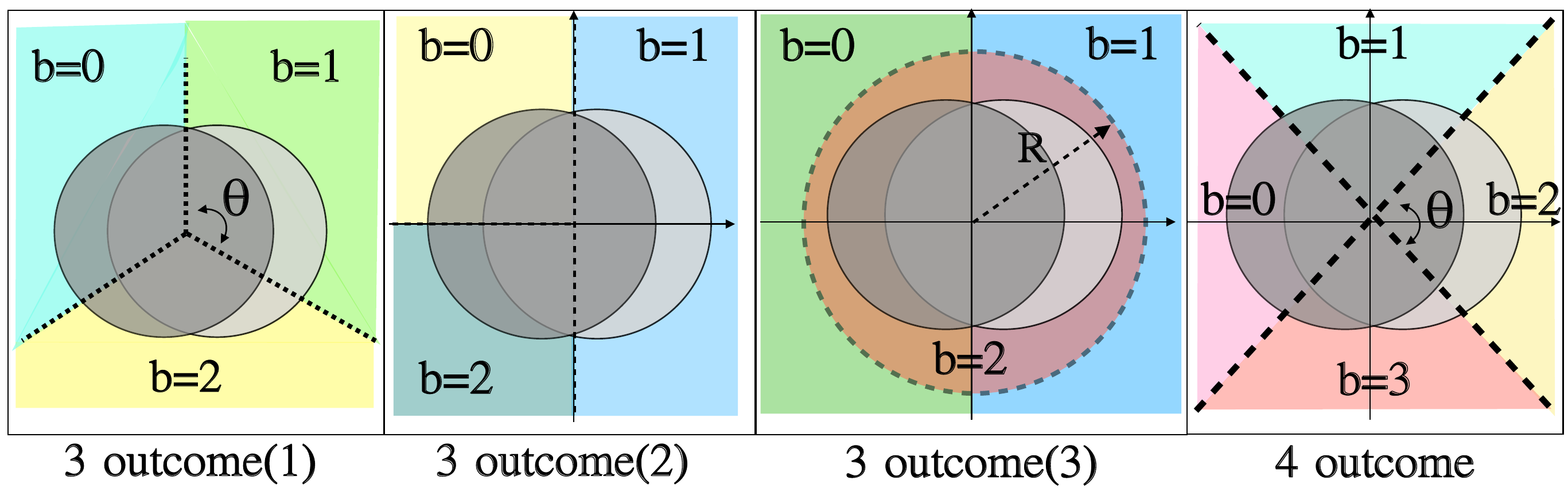}
	\caption{Alternative partitions of the phase space for Heterodyne measurement}
	\label{fig:conf2}
    \end{figure}

    \begin{figure}[h!]
    \includegraphics[width=\linewidth]{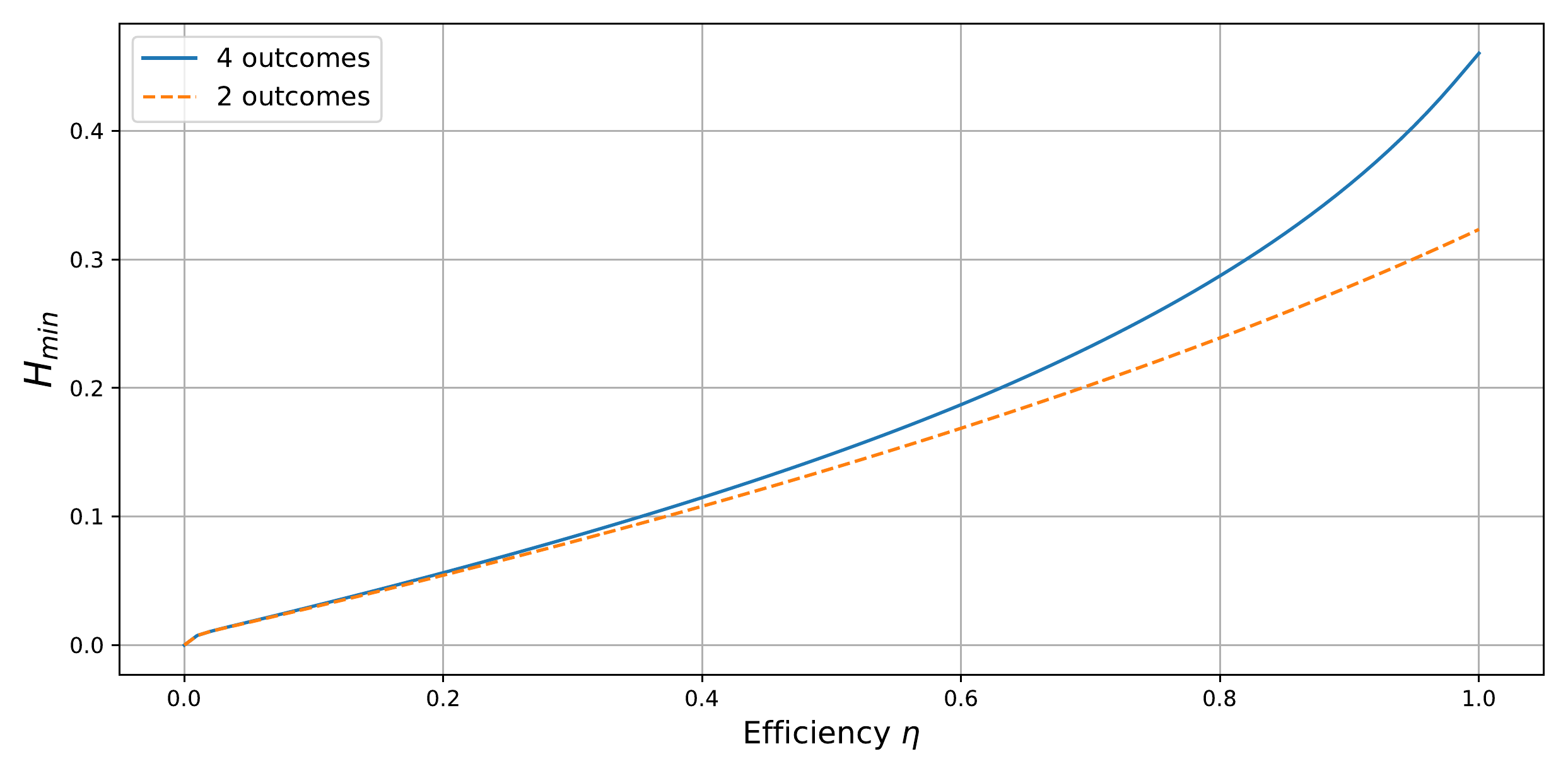}
	\caption{Min-entropy as a function of efficiency $\eta$ for the homodyne detection concerning 2 and 4 outcomes configuration. The mean-photon number $\mu$ and range $L$ are chosen in a way that the min-entropy is maximized.}
	\label{fig:eta1}
    \end{figure}

    \section{Practical considerations }

    The main focus of this work is studying the influence of extending the number of outcomes on Semi-DI QRNG based on an energy bound and homodyne or heterodyne detection. We focused on homodyne and heterodyne detection because they are the most common measurement schemes employed in CV protocols. Moreover, recent experiments ~\cite{Rusca2020, avesani2020}, employed these measurements schemes to implement energy-bounded Semi-DI QRNG protocols. These works could benefit from this analysis, without any modifications to the experimental setup. In fact, the presented results show an enhancement of the certifiable min-entropy with respect to the binary case for ideal detection and no losses. However, we note that in practical implementations the expected improvement is reduced. In fact, additional losses, limited detector's efficiency and excess noise of the receiver apparatus contribute to a reduction of the correlations $p(b|x)$, limiting the advantage of these schemes, as shown in Fig. \ref{fig:eta1}.
 
    We note that, as shown in Fig. \ref{fig:homodyne}, there is almost no improvement when the general inefficiency of the experiment $\eta$ is lower than $12.5\%$. 
    Any experimental realization that would like
    to exploit the advantage of many-outcome configuration should be designed in order to 
    achieve high efficiency.
    
    Although by the homodyne detection higher randomness can be certified with respect to heterodyne detection, the former it is susceptible to errors in the setting of the phase $\phi$ between the signal and the LO.     Indeed, phase errors induces information loss in homodyne detection, whose magnitude depends on the active phase stabilization response time and precision. It is possible to show that 
    a homodyne detection with phase error $\delta\phi$ is equivalent to a homodyne detection with no phase error and efficiency
    $\eta=|\cos(\delta\phi)|$.
    In Fig. \ref{fig:phase} we show the optimal min-entropy for a 4-outcome homodyne detection as a function of the phase error.
    As an example, if the phase error is below $15^\circ$, the min-entropy may fluctuate between $0.47$ and $0.4$. 
    On the other hand, heterodyne detection is robust with respect to phase error as long as the sampling rate is much larger than the phase drift: in the latter case, phase-compensations techniques can be used to track and correct phase fluctuations, with minimal impact on the min-entropy. As described in \cite{avesani2020}, for the heterodyne detection phase drifts can be  compensated via software during the post-processing of the data (see also Fig.\ref{fig:het_hom} c).
    \begin{figure}[h!]
    \includegraphics[width=\linewidth]{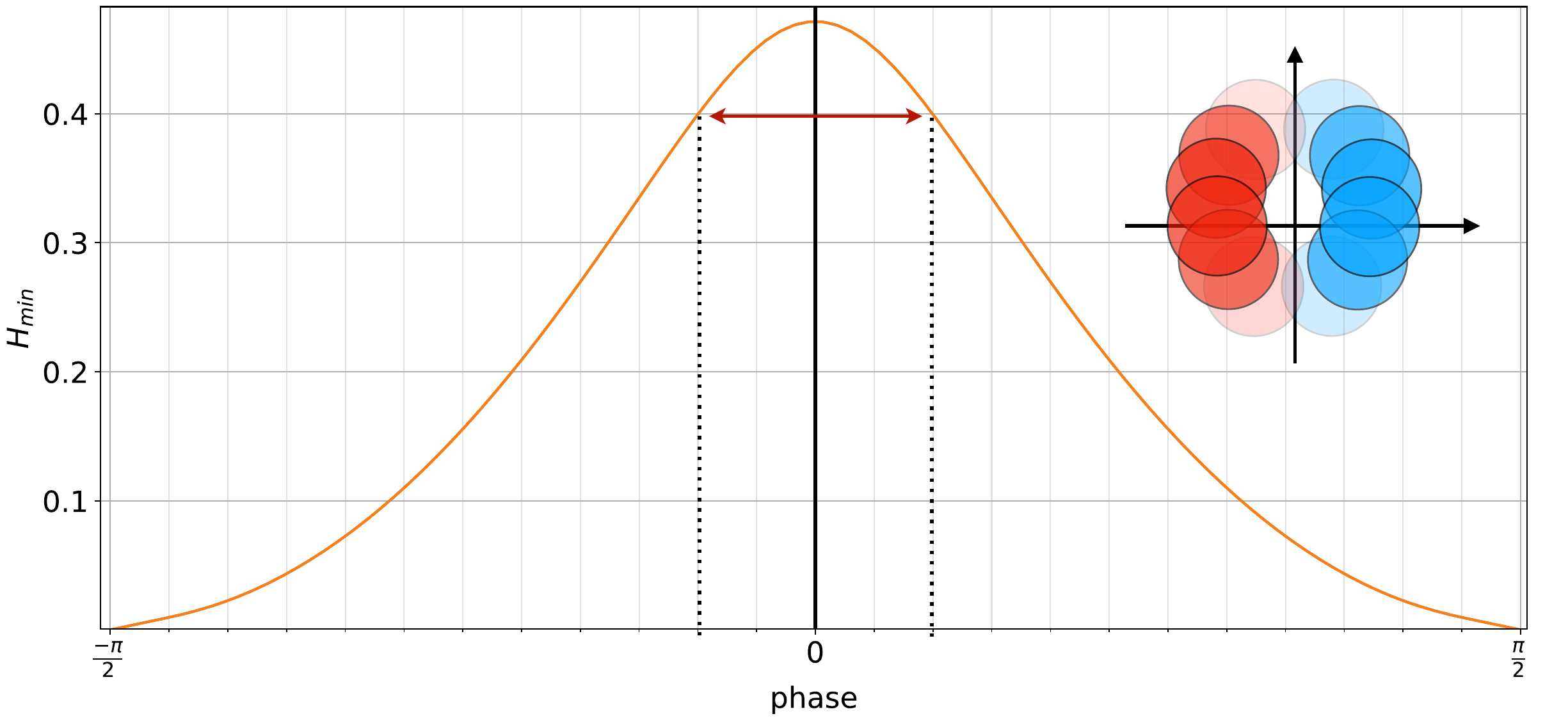}
	\caption{Optimal min-entropy as a function of phase error for 4-outcomes homodyne detection.}
	\label{fig:phase}
    \end{figure}

    \section{Conclusion}
    We have demonstrated a semi-DI QRNG with $d$-outcomes for binary-encoded optical coherent states based on heterodyne or homodyne detection. We compared our results with the binary-outcomes case, and we showed the number of certified random bits improves by increasing the number of outcomes. In this framework, we observed that the homodyne receiver beats the heterodyne receiver in terms of generated randomness. Numerically, we  found an asymptotic upper bound of $0.5$ as the number of random bit per measurement  in the limit of infinite outcomes. 
    
    Moreover, in the heterodyne case we found the partition of the phase space into
    vertical ``strip'' allow an higher generation rate with respect to other configuration (see Fig. \ref{fig:homodyne}). Physically, this could be interpreted by a better discrimination between the two input states
    with the strip configuration compared to other phase-space partitions.
    
     From previous analysis~\cite{Mari2019},
     it is known that the maximum entropy for binary input setup is $\log_2 (3)\simeq1.5849$, while our 
     analysis seems to indicate that with homodyne and heterodyne measurement one can never exceed $0.5$ bit of randomness per measurement.
     We leave for future works the formal proof of the above observation.

    It is worth to note that the improvement is significant for perfect detection efficiency, while it decreases in case of losses. Hence, owning efficient and low-noise detectors is essential for exploiting $d$-outcomes configuration and obtaining higher randomness with respect to the binary-outcome setting. Finally, we illustrated how to apply the $d$-outcome configuration to the experimental data.
    
    \begin{acknowledgments}
    This work was supported by: ``Fondazione Cassa di
    Risparmio di Padova e Rovigo'' with the project QUASAR
    funded within the call ``Ricerca Scientifica di Eccellenza
    2018''; MIUR (Italian Minister for Education) under the initiative ``Departments of Excellence'' (Law 232/2016); EU-H2020 program under the Marie
    Sklodowska Curie action, project QCALL (Grant No. GA 675662).
    \end{acknowledgments}

    \appendix

    \section{Dual SDP}

    In the present section, we report how to dualize the primal form of SDP Eq.(\ref{eq:SDP_primal}). The SDP duality gives an approach to upper bound the optimal value of maximization problems, or a lower bound for minimization problems \cite{boyd2004convex}. The dual SDP has several advantages over the primal version. 
    First, the dual optimization problem returns an upper-bound on the guessing probability, while the primal problem returns a lower-bound. Thus, even if the solver doesn't converge to the exact optimal point, the dual solution will never overestimate the true content of randomness, providing reliable bounds. Secondly, for real-time operation, the dual problem enables to recompute (sub-optimal) bounds without the need of running a full optimization, reducing the resources needed for the entropy estimation.
    Finally, in the dual problem the finite-size effects can be taken into consideration efficiently, thanks to the linear dependance of the $p(b|x)$ in the objective function. Note that in the real experiment, the conditional probabilities $p(b|x)$ are calculated over finite raw data; thus, finite-size effects must be accounted for estimating the bound.

    By using the Lagrangian duality~\cite{boyd2004convex}, with an approach a similar to the one used in \cite{Brask2017}, the dualized SDP can be written as 
    \beq
     P^*_g = \mathop {\min}\limits_{{H^{{\lambda _0},{\lambda _1}}},{\nu _{bx}}}  [- \sum_{x=0,1}\sum_{b=0}^{d-1} {{\nu _{bx}}p(b|x)}] 
    \label{objec}
    \eeq
    subjected to
    \beq
    \begin{aligned}
    &{H^{{\lambda _0},{\lambda _1}}} = {\rm{ }}{({H^{{\lambda _0},{\lambda _1}}})^\dag }, \\
    \sum\limits_x & {{\rho _x}(\frac{1}{2}\sum\limits_{b = 0}^{d-1}\delta_{\lambda_x,b} + {\nu _{bx}})} \\
    &+ {H^{{\lambda _0},{\lambda _1}}} - \frac{1}{2} \tr[{H^{{\lambda _0},{\lambda _1}}}]\openone \le 0
    \end{aligned}
    \label{const}
    \eeq
    where $ H_b^{{\lambda _0},{\lambda _1}}$ are $2\times 2$ Hermitian matrices.
    
    As we can see, the objective function of dual SDP is a linear function of the conditional probability distribution $p(b|x)$, and the these are not appearing in the constraints. Hence, after solving the dual SDP one time and obtaining a valid set of parameters $\nu_{bx}^*$, it is possible to obtain a (sub-optimal) bound for a new set of experimental probabilities $p(b|x)$, by evaluating the objective linear function with the set of parameters $\nu_{bx}^*$.
    This estimation doesn't require the full optimization of the SDP, which can be slow and could limit the rate in real-time operation.
    A similar appproach is not possible with the primal version that needs to run full optimization of the SDP for every new set of $p(b|x)$.
    \\
    \\

\bibliography{bibliography}

\end{document}